\documentclass[a4paper, 10pt]{article}
\usepackage[margin=2.5cm]{geometry}

\usepackage{graphicx, subfig, multicol, multirow, seqsplit}
\graphicspath{{_figures/}}
\usepackage[table]{xcolor}
\usepackage{cite}
\usepackage{booktabs} 
\usepackage{hyperref}	
\hypersetup{colorlinks=true, citecolor=blue, linkcolor=blue, urlcolor=blue,pdfpagemode=UseNone}
\usepackage{amssymb, amsmath}

\usepackage{setspace}
\usepackage{placeins} 
\usepackage{adjustbox}

\usepackage{array}
\newcolumntype{L}[1]{>{\raggedright\let\newline\\\arraybackslash\hspace{0pt}}m{#1}}
\newcolumntype{C}[1]{>{\centering\let\newline\\\arraybackslash\hspace{0pt}}m{#1}}
\newcolumntype{R}[1]{>{\raggedleft\let\newline\\\arraybackslash\hspace{0pt}}m{#1}}


\usepackage[labelfont=bf,labelsep=period,justification=justified, singlelinecheck=false]{caption}

\makeatletter 		\renewcommand{\@biblabel}[1]{\quad#1.} 			\makeatother
\date{}
\newcommand{\modelname}[1]{\texttt{\seqsplit{#1}}}
\newcommand{\mlp}{\modelname{MLP}}
\newcommand{\deepmlp}{\modelname{DeepMLP}}
\newcommand{\deepresmlp}{\modelname{DeepResMLP}}
\newcommand{\qwenthree}{\modelname{Qwen3-Embed-0.6B}}
\newcommand{\qwentwo}{\modelname{gte-Qwen2-1.5B-instruct}}
\newcommand{\gtemultilingual}{\modelname{gte-multilingual-base}}
\newcommand{\snowflakel}{\modelname{snowflake-arctic-embed-l-v2.0}}
\newcommand{\snowflakem}{\modelname{snowflake-arctic-embed-m-v2.0}}
\newcommand{\nomic}{\modelname{nomic-embed-text-v1.5}}

\title{\textbf{Federated Learning for ICD Classification with Lightweight Models and Pretrained Embeddings}}\label{faiicd10}
\author{Binbin XU, Gérard DRAY}
\date{\normalsize EuroMov Digital Health in Motion, Univ. Montpellier, IMT Mines Ales, France\\
\texttt{binbin.xu@mines-ales.fr}, \texttt{gerard.dray@mines-ales.fr}}

\begin{document}
\maketitle

\begin{abstract}
This study investigates the feasibility and performance of federated learning (FL) for multi-label ICD code classification using clinical notes from the MIMIC-IV dataset. Unlike previous approaches that rely on centralized training or fine-tuned large language models, we propose a lightweight and scalable pipeline combining frozen text embeddings with simple multilayer perceptron (MLP) classifiers. This design offers a privacy-preserving and deployment-efficient alternative for clinical NLP applications, particularly suited to distributed healthcare settings.
Extensive experiments across both centralized and federated configurations were conducted, testing six publicly available embedding models from Massive Text Embedding Benchmark leaderboard and three MLP classifier architectures under two medical coding (ICD-9 and ICD-10). Additionally, ablation studies over ten random stratified splits assess performance stability. Results show that embedding quality substantially outweighs classifier complexity in determining predictive performance, and that federated learning can closely match centralized results in idealized conditions.
While the models are orders of magnitude smaller than state-of-the-art architectures and achieved competitive micro and macro F1 scores, limitations remain including the lack of end-to-end training and the simplified FL assumptions. Nevertheless, this work demonstrates a viable way toward scalable, privacy-conscious medical coding systems and offers a step toward for future research into federated, domain-adaptive clinical AI.
\end{abstract}

\section{Introduction}\label{sec:intro}

Modern healthcare systems generate large amounts of unstructured clinical data, particularly in the form of free-text medical notes. Accurately classifying these records using standardized coding systems like the International Classification of Diseases \cite{WorldHealthOrganization2009International} (ICD-9 or ICD-10) is essential for epidemiological investigation, and clinical research. 
However, manual coding is time-consuming, error-prone, and increasingly unsustainable given the scale and complexity of electronic health records (EHRs). While machine learning (ML), especially natural language processing (NLP) with large language models (LLMs), offers significant potential in automating this task, the practical deployment of such models in healthcare is constrained by regulatory and ethical concerns surrounding patient data privacy. Traditional ML approaches rely on centralized training, which requires aggregating sensitive health records which is often impossible due to data protection laws such as the General Data Protection Regulation (GDPR), institutional / ethical barriers, and patient trust concerns. Consequently, data remains fragmented across hospitals and health systems, limiting the development of robust, generalizable models. Federated learning has emerged as a potential paradigm for overcoming these barriers by enabling collaborative model training without direct data sharing. Yet despite increasing attention, FL remains underutilized in real-world healthcare NLP applications, particularly in complex multi-label classification tasks like ICD coding.

Automated ICD classification has been studied with a wide range of machine learning and natural language processing techniques. Early approaches focused on linear models such as logistic regression and support vector machines (SVMs), leveraging simple text features like \texttt{bag-of-words} and \texttt{tf-idf} weights. These models often treated each ICD code as an independent binary classification problem, with modest gains achieved through hierarchical structuring to reflect the ICD ontology \cite{Perotte2013Diagnosis, Xie2019EHR}. 
As deep learning evolved, a second wave of approaches emerged, including sequential and convolutional architectures such as CNNs, BiGRUs, and LSTMs to better model the semantic structure of clinical documents \cite{Mullenbach2018Explainable, Baumel2018Multi,Li2019Automated}. The research focus was then shifted to more sophisticated techniques, like attention mechanisms (e.g., CAML, DR-CAML) and label-aware embeddings (e.g., LEAM), which improved the processing of multi-label outputs and rare codes \cite{Mullenbach2018Explainable, Wang2018Joint}.

To further integrate domain knowledge, some systems incorporated external ontologies and graph-based structures, for example MSATT-KG which combined multi-scale CNNs with graph neural networks to model code relationships \cite{Xie2019EHR}. Other works took rule-augmented or expert-driven approach, such as feature selection pipelines in autopsy analysis \cite{Mujtaba2017Automatic} and hybrid models for disease classification in death certificates \cite{Koopman2015Automatic}. Even language-specific adaptations, like SVM pipelines for Bulgarian discharge summaries, attempted to bridge terminological gaps between clinical text and ICD codebooks as early as 2011 \cite{Boytcheva2011Automatic}.

Despite these incremental improvements, most of the earlier systems remained narrowly tailored and often limited in generalizability. They required generally extensive manual tuning, domain-specific ontologies, or handcrafted feature engineering which restricted in fact the portability across healthcare settings, or code systems. Moreover, due to parameter space limitations, these architectures struggled to capture contextual semantics or exploit unlabeled clinical data, and were not very suitable for transfer learning or zero-shot inference. These limitations made them inadequate for modern and large scale ICD classification tasks, which require robustness to domain drift, label imbalance, and complex clinical environment.

The emergence of large language models (LLMs), particularly transformer-based architectures pretrained on large dataset, showed a methodological breakthrough. LLMs such as BERT and GPT architectures are able to provide contextualized text understanding, few-shot learning, and generalizability without task-specific feature engineering. In clinical coding, early studies demonstrated state of the art (SOTA) results using fine-tuned LLMs like PubMedBERT, ClinicalBERT, or RoBERTa for ICD-10 assignment \cite{Chen2022Training} or other Domain-Specific Pretraining approach \cite{Huang2022PLM}.

In our prior work in 2019, we trained a GPT-2 model from scratch on emergency department notes in French for injury classification \cite{Xu2020Pre-Training}. This model significantly reduced the volume of labeled data needed to achieve high accuracy, confirming the sample efficiency benefits of generative pretraining in a real-world, non-English dataset. However, due to legal barriers to data sharing, this system could not be extended across institutions. And to build a more complete system capable of classifying a wider range of ICD codes, data access-not model capacity-became the fundamental bottleneck. 

While the potential of LLMs for clinical coding is now well recognized, few studies have examined whether such models can be trained in federated learning (FL) framework, in line with privacy and legal requirements. A notable and promising work by Chen et al. \cite{Chen2022Training}, applied federated learning to train a BERT-based model across three hospitals for ICD-10 coding. Although this model outperformed locally trained baselines, its scope was limited by the number of sites. Meanwhile, other FL studies in healthcare-such as those by Budrionis et al. \cite{Budrionis2021Benchmarking}, Mondrejevski et al. \cite{Mondrejevski2022FLICU}, and Horvath et al. \cite{Horvath2023Exploratory} - explored federated workflows using simpler architectures (e.g., shallow neural networks, LSTMs, or CNNs), focusing more the classification tasks like mortality prediction rather than multi-label code assignment from complex clinical texts.

In this study, we propose a hybrid approach that leverages the semantic strength of large language models while addressing core challenges of federated learning in healthcare: privacy protection, standardized representation, and deployment scalability. Rather than fine-tuning the whole transformer models across institutions - a process that is compute-intensive, communication-heavy, and privacy-sensitive - we adopt a modular architecture that decouples representation from classification. Specifically, we use several top-performing, open-source embedding models (e.g., Qwen3 0.6B, Qwen2-1.5B and others from the MTEB leaderboard), licensed under Apache 2.0, to encode clinical notes into dense vector representations. These embeddings are computed offline and locally, allowing each hospital to deploy the same standardized model within its infrastructure without transferring any raw data to the external server for centralized training.

This offline encoding paradigm not only enhance data sovereignty but also addresses a major weakness in federated NLP: the lack of consistent and standardized feature spaces across nodes. Because embeddings are fixed and locally normalized, and because their dimensionality (768–1536) precludes any practical text reconstruction, this setup provides a strong privacy-preserving abstraction layer. The federated training then focuses only on lightweight classification heads (MLP, DeepMLP, DeepResMLP), which are trained collaboratively without any access to the original clinical text. Though these architectures are relatively simple, they operate on semantically enriched, standardized inputs-allowing the system to capture complex diagnostic relationships without centralizing sensitive information. This study evaluates federated multi-label classification for ICD-9 and ICD-10 codes using LLM-derived embeddings, balancing state-of-the-art performance with real-world deployment constraints.

\section{Method}\label{sec:datask}

\begin{figure}[!htb]
    \centering
    \includegraphics[width=\linewidth]{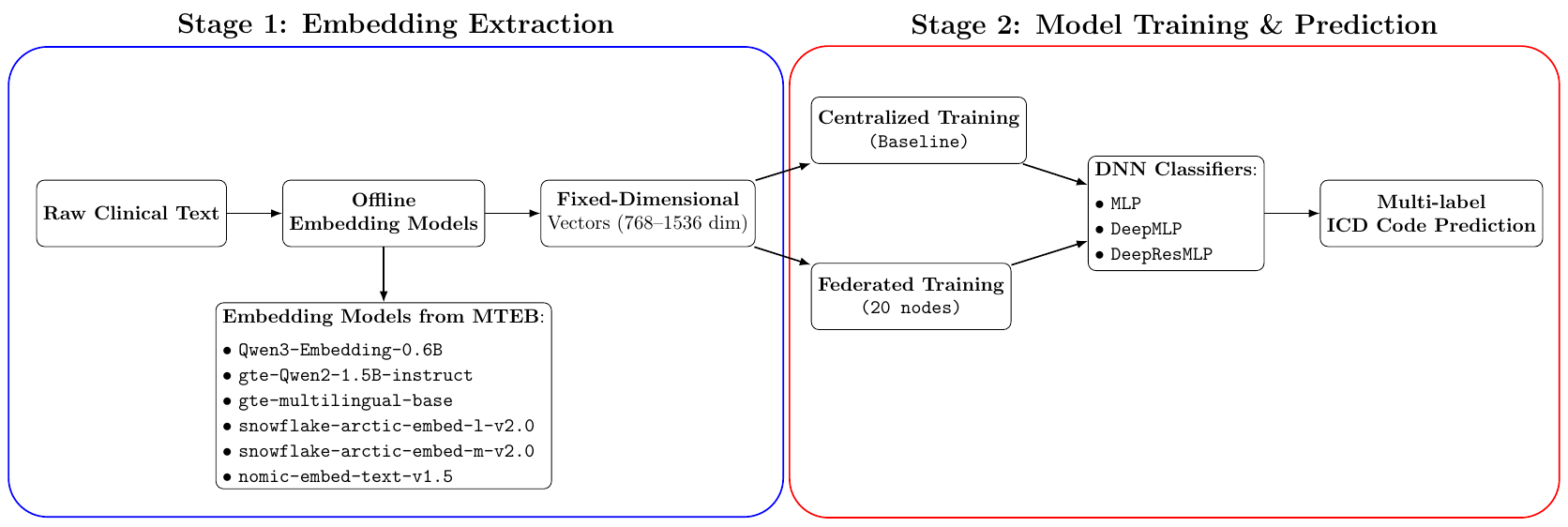}
    \caption{Overview of the proposed pipeline. Raw clinical text is encoded into dense embeddings using open-source models, followed by centralized or federated training of lightweight classifiers for multi-label ICD coding.}
    \label{fig:overview}
\end{figure}

To structure the learning pipeline and enforce strict modularity between semantic encoding and classification, our approach is divided into two stages: (1) offline embedding extraction using open-access large language models, and (2) classifier training via centralized or federated learning.
Figure~\ref{fig:overview} illustrates the two-stage architecture adopted in this work. In the first stage, clinical notes are encoded using six SOTA open-source models from the MTEB leaderboard\footnote{\url{https://huggingface.co/spaces/mteb/leaderboard}}, all deployed offline to ensure data privacy and reproducibility. This step yields standardized, fixed-dimensional embeddings (upon models, respectively 768, 1024, 1536) without transferring any raw text outside the local site if deployed in different sites. The second stage consists of model training using either centralized (as baseline study) or federated learning strategy across 20 nodes, with identical training and evaluation datasets and embedding protocols. Three classifier architectures-MLP, DeepMLP, and DeepResMLP-are trained to perform multi-label ICD-9 and ICD-10 code classification. This decoupled design isolates the contribution of the training architecture while preserving semantic representation from the LLM embeddings.

\subsection{Data}

Modern NLP approaches, particularly those involving deep learning, generally require access to large-scale textual data. While general-domain tasks benefit from the broad linguistic and semantic knowledge captured by SOTA LLMs, this transferability breaks down in specialized domains such as healthcare. Medical texts exhibit domain-specific vocabulary, structural irregularities, and diagnostic nuance that are in general rarely represented or totally absent in public pretraining corpora. Consequently, even high-performing LLMs often show reduced performance on clinical tasks compared to general-purpose NLP benchmarks.

Importantly, large-scale clinical text does exist within individual healthcare institutions. In prior work \cite{Xu2020Pre-Training}, we trained a GPT-2 model from scratch using over 300\,000 emergency department notes in French from a single French hospital. This volume represented only a subset of that hospital's total EHR corpus, and it's an even smaller fraction of the national record system. However, despite the technical success of that approach, legal and privacy constraints made it infeasible to expand model training beyond a single hospital. This bottleneck illustrates a critical challenge in clinical NLP: the paradox of abundant but isolated data.

To simulate a more realistic multi-institutional setting while avoiding regulatory limitations, we turn to the Medical Information Mart for Intensive Care (MIMIC) database. In this study, we use the MIMIC-IV-Note corpus \cite{Johnson2023MIMIC}, the largest publicly accessible source of clinical free-text. It contains 331\,794 de-identified discharge summaries from 145\,915 patient stays, in addition to 2\,321\,355 radiology reports from 237\,427 patients. All patient identifiers are removed, and access is granted via PhysioNet \cite{Goldberger2000PhysioBank} under a data use agreement that ensures ethical handling. This dataset serves as a proxy for large-scale institutional data while remaining compliant with open research standards.

For our study, we focus exclusively on discharge summaries labeled with either ICD-9 or ICD-10 codes. To ensure a meaningful and balanced distribution for federated learning across clients, we impose the following constraint: each label must be associated with at least 10 samples per client, across 20 clients. This translates to a minimum of 200 samples per label in the overall dataset. After filtering under this criterion, we retain:
\begin{itemize}
    \item 121\,344 notes with 1\,085 unique ICD-10 codes, and
    \item 207\,817 notes with 1\,267 unique ICD-9 codes.
\end{itemize}
In total, 329\,161 discharge summaries are used in this study. These are evenly split across the centralized and federated training settings to ensure a controlled performance comparison.

\subsection{Embedding and Representation}

Large language models have demonstrated significant performance and generalizability across a wide range of NLP tasks, often achieving SOTA results in zero-shot or few-shot settings. Their ability to capture rich semantic structures and domain-independent language representations makes them very useful for downstream tasks in specialized fields such as healthcare, where labeled data is often scarce or siloed.

Among the various modalities of LLM usage, embedding models-transformers trained to produce fixed-length vector representations of text-have become foundational in modern NLP. Tasks such as information retrieval, sentence / document classification, semantic search, and clustering are now routinely powered by dense vector embeddings. 

Beyond their semantic utility, embedding models provide also a crucial secondary privacy layer in clinical NLP. Unlike raw text, dense vector embeddings are non-invertible, lossy representations of the input: the transformation is many-to-one, nonlinear, and not uniquely decodable. While inversion methods, such as supervised decoders or autoencoders, could sometimes approximate paraphrases, they cannot reconstruct the exact original text, particularly when the embedding model is not jointly trained with a decoder. Transformer encoders like BERT or Qwen, which we use here, compress meaning into high-dimensional space (768–1536 dimensions) in a way that drops word-level and positional detail. 

In this study, we exploit this property deliberately: embedding is performed offline and locally, and the resulting vectors serve as a standardized, privacy-preserving input for downstream single-label or multi-label classification. This approach complements federated learning by ensuring that no raw text or token-level identifiers ever leave the local node, even in pre-encoding stages. One core aim of this study is to evaluate how well this two-layer privacy strategy using offline embedding plus federated classification can scale up to large, multi-label diagnostic coding tasks without sacrificing model performance.

The Massive Text Embedding Benchmark (MTEB) \cite{Muennighoff2022MTEB} has emerged as the primary evaluation suite for such models, offering a standardized, multi-task leaderboard covering dozens of NLP scenarios. As of writing, the MTEB leaderboard ranks 269 embedding models, while over 1\,900 models on HuggingFace are tagged with MTEB-relevant metadata. To identify embedding models suitable for clinical use under federated constraints, we applied the following filters:
\begin{itemize}
    \item Free and open-source licenses: models must be under the Apache 2.0 or MIT license;
    \item Long input context: models must support input lengths of at least 8\,000 tokens, since clinical documents (e.g., discharge notes) are often lengthy;
    \item Compact size: parameter count under 2 billion, suitable for 16GB VRAM GPU or less, to ensure local deployability in real-world hospital infrastructure;
    \item High semantic quality: ranking in the top 100 on the MTEB leaderboard.
\end{itemize}

\begin{table}[!htb]
\setlength{\tabcolsep}{5pt}
\small
  \centering
  \caption{Embedding models under Apache 2.0 license, from HuggingFace MTEB Leaderboard, access on {June 10\textsuperscript{th}, 2025} }
    \begin{tabular}{c|lcccc}
    \toprule
    \textbf{Rank} & \multicolumn{1}{c}{\textbf{Model}} & \textbf{N.Param} & \textbf{EmbedDim} & \textbf{MaxTokens} & \textbf{MeanTaskScore}  \\ \midrule
    4     & \qwenthree \cite{Zhang2025Qwen3} & 595M  & 1024  & 32768 & 64.34  \\
    14    & \qwentwo \cite{Li2023Towards} & 1.5B    &  1536  & 32768 & 59.45  \\
    25    & \gtemultilingual \cite{Zhang2024mGTE} & 305M  & 768   & 8192  & 58.24  \\
    34    & \snowflakel \cite{Yu2024Arctic} & 568M  & 1024  & 8192  & 57.03  \\
    40    & \snowflakem \cite{Yu2024Arctic} & 305M  & 768   & 8192  & 53.70   \\
    77    & \nomic \cite{Nussbaum2025Nomic} & 137M  & 768   & 8192  & 44.10  \\ \bottomrule
    \end{tabular}%
  \label{tab:embedmdlsapache2}%
\end{table}%

This filtering yielded six models, all of which are used in our study as shown in Table \ref{tab:embedmdlsapache2}. These models span a range of training objectives (contrastive, retrieval-based, instruction-aligned), architectures and design principles, but all meet our criteria for license compatibility, scalability and semantic quality. 
\qwenthree\, \cite{Zhang2025Qwen3, Yang2025Qwen3} and \qwentwo\, \cite{Li2023Towards, Yang2024Qwen2} are from the Qwen family and support long-context and instruction-aligned input when necessary. Model \gtemultilingual\, \cite{Zhang2024mGTE} offers compact multilingual embeddings suitable for low-resource environments. Model \texttt{\seqsplit{snowflake-arctic-embed-l/m-v2.0}} \cite{Yu2024Arctic} prioritize retrieval accuracy using multi-scale contrastive learning. Model \nomic\, \cite{Nussbaum2025Nomic} is optimized for multilingual semantic search. This diversity allows us to assess whether federated training remains robust across heterogeneous embedding backbones. These models have also the multilingual capabilities which are especially important for heterogeneous clinical environments.

Although the MTEB entry lists the \qwentwo\, model with an embedding dimension of 8960, all sources (e.g., the Hugging Face model card) report its actual output vector dimension as 1536. In this study, the actual output was indeed 1536, as verified through local inference.

All embedding models are applied offline, producing dense vector representations of dimension 768 to 1536 depending on the model. The dense vectors are later used as inputs to centralized or federated classifier training. This design supports both uniform representation across nodes and privacy-by-design policies, aligning with real-world clinical deployment requirements. The upper threshold of number of token (\texttt{max\_length}) in all embedding process is set to 8192, representing  $0.21\%-0.35\%$ for ICD-9 and $0.55\%-0.80\%$ for ICD-10 data which is negligible and does not affect the training and evaluation. 

\begin{figure}[!htb]
    \centering
    \includegraphics[width=\linewidth]{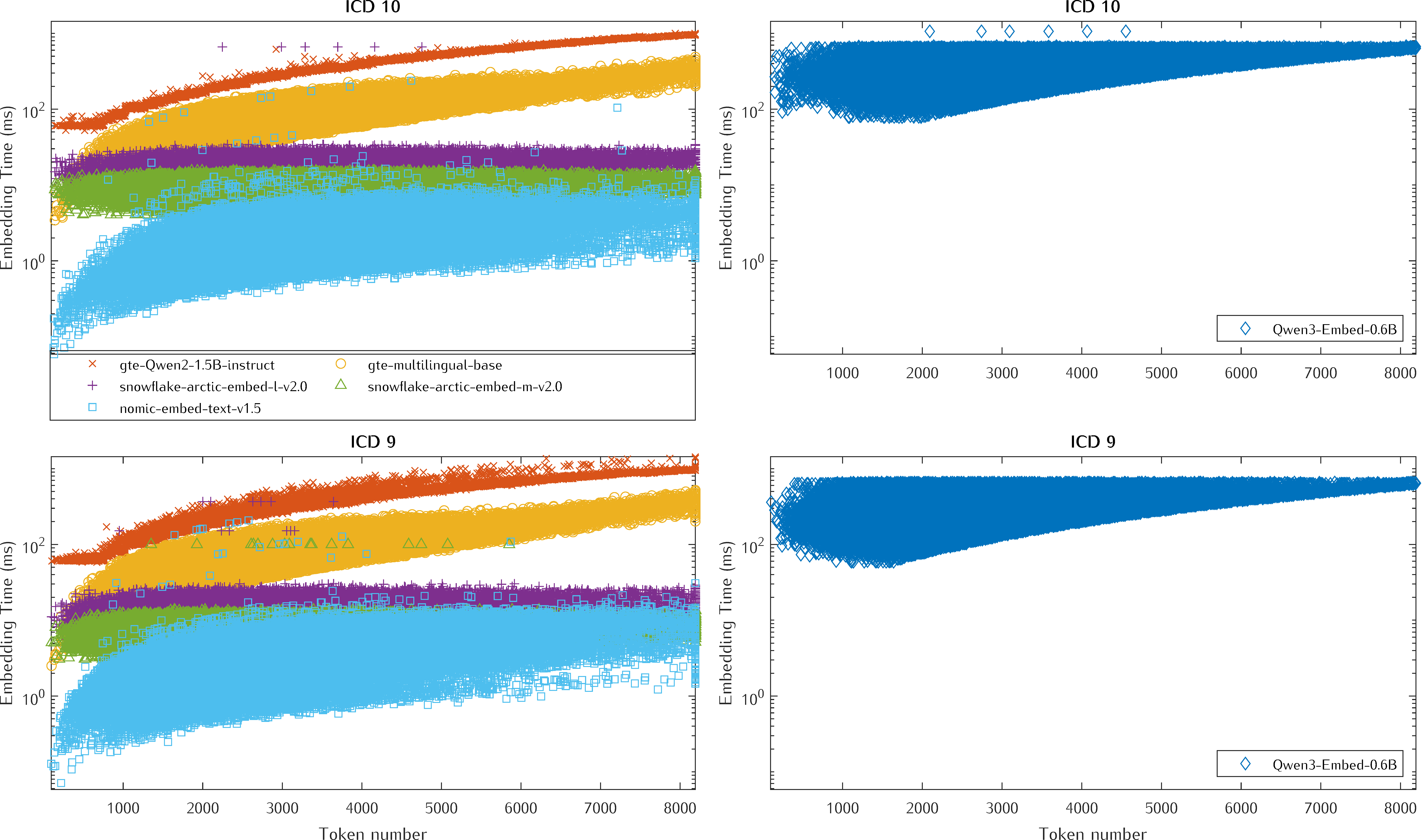}
    \caption{Embedding inference time vs. input token length for six selected models on both ICD-9 and ICD-10 datasets. Embeddings were computed offline. Qwen3-Embedding-0.6B is shown separately due to being run on a different GPU (RTX A5500), while the others were run on an RTX A4000. Note that inference time scales with token length, but remains tractable for all models under typical discharge note lengths.}
    \label{fig:embedtime}
    \vspace{0.5em}
\footnotesize
\setlength{\tabcolsep}{3pt}
    \begin{tabular}{l|C{2cm}C{3cm}C{2cm}C{3cm}}
    \toprule
          & \multicolumn{2}{c}{ICD-9} & \multicolumn{2}{c}{ICD-10} \\ \midrule
    Embedding & Total Count & Count (Tokens==8192) & Total Count & Count (Tokens==8192) \\ \midrule
    \qwenthree & 207\,817 & 737   & 121\,344 & 974 \\
    \qwentwo & 207\,817 & 737   & 121\,344 & 974 \\
    \gtemultilingual & 207\,817 & 486   & 121\,344 & 701 \\
    \snowflakel & 207\,817 & 486   & 121\,344 & 701 \\
    \snowflakem & 207\,817 & 486   & 121\,344 & 701 \\
    \nomic & 207\,817 & 439   & 121\,344 & 665 \\
    \bottomrule
    \end{tabular}%
\end{figure}

Figure~\ref{fig:embedtime} shows the relationship between input token length and inference time (in milliseconds) for the six selected embedding models, evaluated separately on discharge notes labeled with ICD-9 and ICD-10 codes. As expected, inference time generally increases with token count, following an approximately linear or sublinear trend for most models.

Among the five models evaluated on the RTX A4000 GPU, three of them (\nomic, \snowflakel\, and \snowflakem) demonstrate consistently fast inference, with nearly all inputs embedded in under 40ms, regardless of length up to 8000 tokens. These models are highly efficient and well-suited for deployment on standard GPU hardware. \gtemultilingual\, shows moderately higher latency, scaling up to 400–500ms for longer documents. \qwentwo\,, while among the strongest semantically, incurs relatively higher computational cost, with embedding time reaching 900–1000ms for full-length samples. \qwenthree\, is shown separately, as it was encoded on a RTX A5500 GPU due to suboptimal performance on A4000 hardware. Its profile remains efficient (under 600-700ms par document), and though direct comparisons should be interpreted with caution, its scaling trend is broadly similar to other models in the study. To summarize, though a small number of long clinical documents result in inference times approaching 600–1000ms, the vast majority of samples fall well below 200ms. 

This further reinforces the practicality of our approach: even for thousands-token documents, the embedding step can be efficiently batched and parallelized offline. Importantly, this embedding step is fully decoupled from training and performed offline, its runtime has no impact on subsequent model training. This separation is a key architectural decision: it enables privacy-preserving pre-processing and allows even heavier models to be used when semantic richness outweighs speed constraints. These results reinforce the feasibility of our approach under realistic deployment constraints in federated settings.

\subsection{Classifier Models}

Given that embedding models in this study already transform free-text notes into dense semantic representations (of 768–1536 dimensions), the role of downstream classifiers becomes relatively lightweight: translating semantically rich embeddings into multi-label outputs. 

The semantic abstraction, typically processed by deep or recurrent networks, is now front-loaded into the embedding stage. As a result, deploying complex transformer-based or recurrent architectures as classifiers would yield limited benefit at disproportionately higher computational cost, particularly in a privacy-preserving, resource-constrained federated setting. Hence, we explored a modular decoupled pipeline and focus on the classification heads, exploring three feedforward variants: MLP, DeepMLP and DeepResMLP. These models are designed to balance expressive capacity with training stability and communication efficiency in a federated configuration. 

\begin{figure}[!htb]
    \subfloat[MLP]{
        \includegraphics[height=3cm]{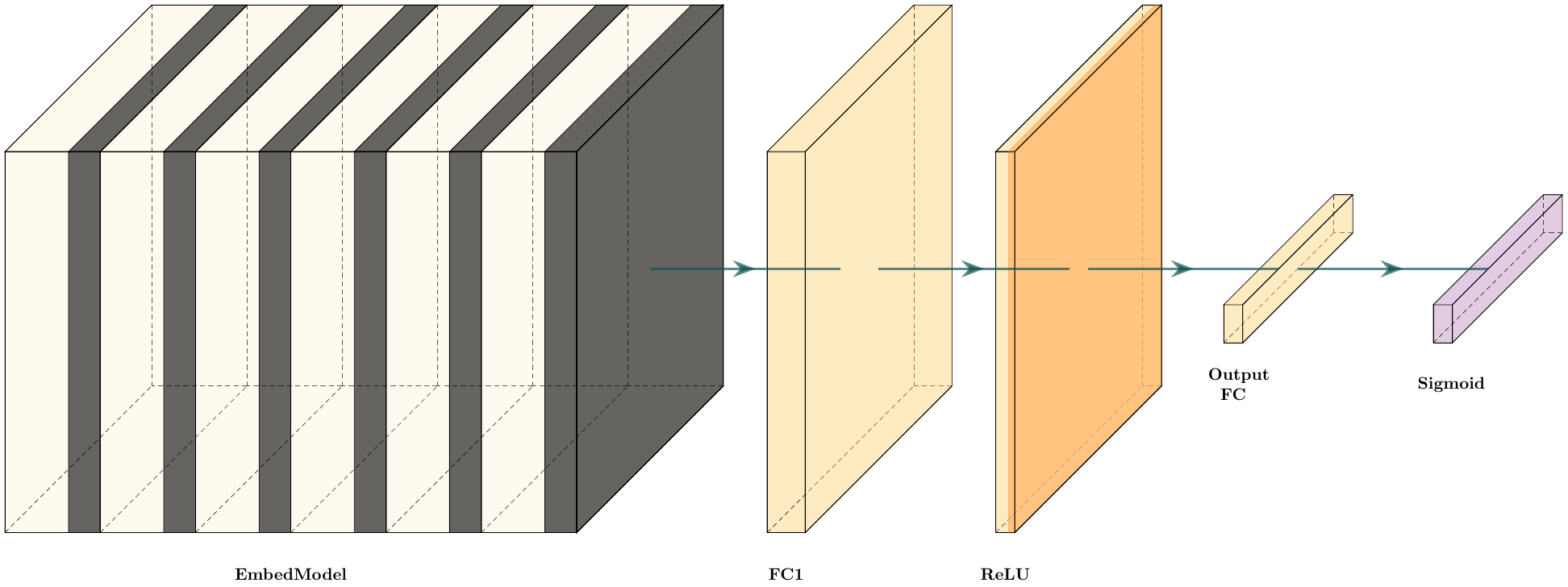}
        \label{fig:mdl_mlp}
    }\\[1em]
    \subfloat[DeepMLP]{
        \includegraphics[height=3cm]{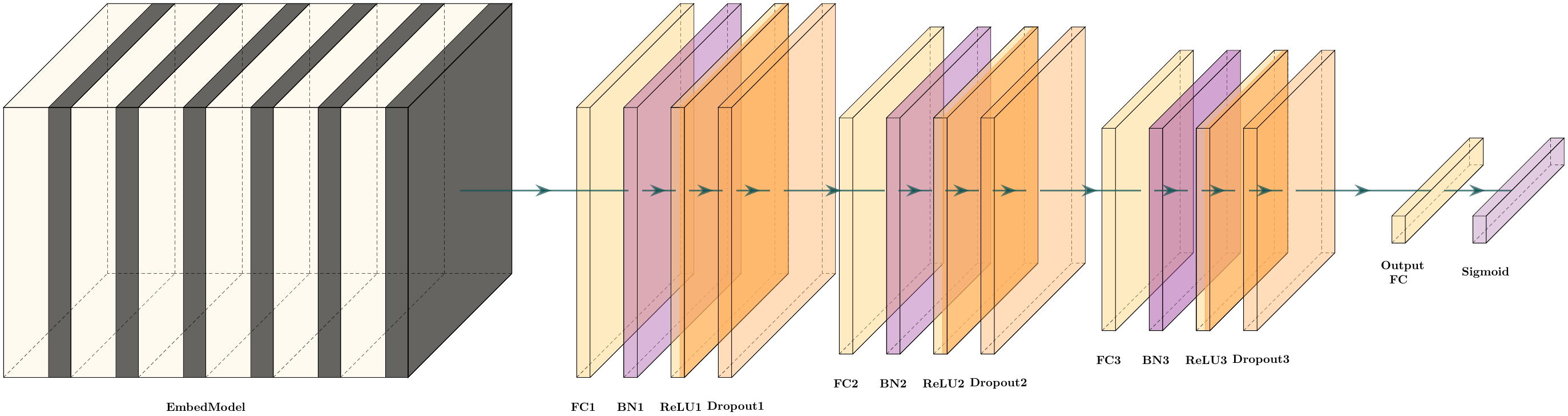}
        \label{fig:mdl_deepmlp}
    }\\[1em]
    \subfloat[DeepResMLP]{
        \includegraphics[height=3cm]{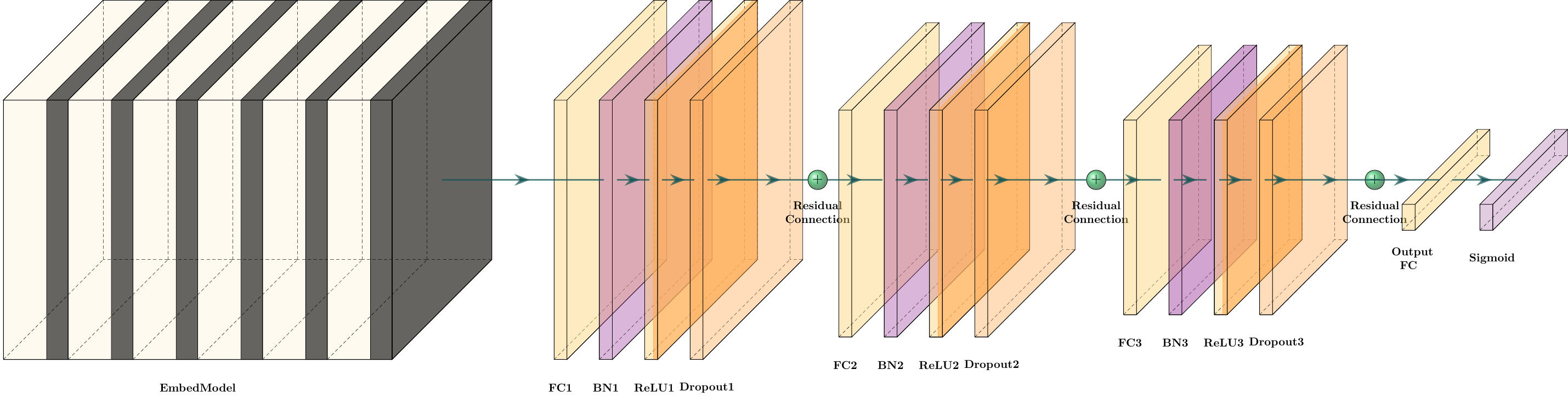}
        \label{fig:mdl_deepresmlp}
    }
    \caption{Models architectural overview. The wide left blocks represent the precomputed embedding vectors.}
    \label{fig:models}
\end{figure}

Figure~\ref{fig:models} provides an architectural overview of the three classification models used in this study: (a) MLP, (b) DeepMLP and (c) DeepResMLP. The embedding model is shown as a wide, front-end backbone block, indicating the heavy lifting performed in semantic representation. All three models share the same input structure: dense embeddings (768–1536 dimensions) generated from the offline encoder stage. The MLP baseline uses a single fully connected (FC) hidden layer followed by a sigmoid output. DeepMLP adds depth with three stacked FC layers interleaved with batch normalization, ReLU activations and dropout for regularization. Finally, DeepResMLP incorporates residual connections across layers to facilitate gradient flow and mitigate degradation in deeper networks.

All models are flexible in input dimension, allowing seamless integration with any embedding model used. This modular  and lightweight classification setup allowed efficient training while keeping model updates compact which is essential for federated learning scenarios. It also aligns with one of the core aims of this study: to demonstrate that with high-quality embedding representations, relatively simple models can achieve strong performance on complex multi-label tasks, even under strict privacy constraints.

Even the largest configuration (DeepResMLP with 1536-dim input and 1267 output labels) contains under 7.1 million parameters (Table~\ref{tbl:paramest}), which is orders of magnitude smaller than typical transformer models-underscoring the feasibility of deploying and training these classifiers in resource-constrained or federated settings. As listed in the work \cite{Edin2023Automated}, CAML \cite{Mullenbach2018Explainable} has 6.1M parameter; the Bi-GRU \cite{Mullenbach2018Explainable}, 9.9M; MultiResCNN \cite{Li2020ICD}, 11.9M; LAAT \cite{Vu2021label}, 21.9M and PLM-ICD \cite{Huang2022PLM}, 138.8M. 

\begin{table}[h]
\small
\centering
\caption{Estimated number of parameters for each model across input/output dimension configurations.}
\label{tbl:paramest}
\begin{tabular}{l|cc|cc|cc}
\toprule
\textbf{Model} & \multicolumn{2}{c|}{\textbf{Input: 768}} & \multicolumn{2}{c|}{\textbf{Input: 1024}} & \multicolumn{2}{c}{\textbf{Input: 1536}} \\
               & \textbf{Out:1085} & \textbf{Out:1267}     & \textbf{Out:1085} & \textbf{Out:1267}       & \textbf{Out:1085} & \textbf{Out:1267} \\
\midrule
MLP            & 1.6M         & 1.8M            & 1.9M         & 2.0M              & 2.4M         & 2.6M         \\
DeepMLP        & 2.1M         & 2.2M            & 2.4M         & 2.6M              & 3.1M         & 3.2M         \\
DeepResMLP     & 4.3M         & 4.3M            & 5.1M         & 5.3M              & 6.9M         & 7.0M         \\
\bottomrule
\end{tabular}
\end{table}

\subsection{Federated Learning Framework}\label{sec:fedai}

To implement federated training of this multi-label classification task, we evaluated several open-source FL libraries, with a focus on simplicity, cross-platform support and performance:
\begin{itemize}
    \item Flower \cite{Beutel2020Flower}, is a lightweight, framework-independent FL platform designed for both research and real-world deployment. It supports heterogeneous devices, different machine learning backends (PyTorch, TensorFlow, even directly NumPy) and can simulate large-scale federations easily. Importantly, it is cross-platform and has an active community and production use cases. \footnote{\url{https://github.com/adap/flower}}

    \item PySyft \cite{Ziller2021PySyft}, specializes in privacy-enhancing such as secure multi-party computation and differential privacy. Though it's very powerful for privacy-preserving computation, it shifts focus away from efficient federated model training and requires heavier abstractions which is often difficult to setup in hospitals. \footnote{\url{https://github.com/OpenMined/PySyft}}
    
    \item TensorFlow Federated, built on Google's TensorFlow, is one of the earliest FL framework and has been extensively evaluated. However it lacks official Windows support (users must rely on Linux or WSL) and tight integration with PyTorch or heterogeneous hardware, making it be inconvenient for real-world applications when most hospitals or institutes still use largely Windows environments. 

    \item FedLab \cite{Zeng2023FedLab}, is a PyTorch-native framework focused on simulation with efficient communication and flexibility. While promising, it is still in early stages with limited ecosystem maturity compared to Flower. \footnote{\url{https://github.com/SMILELab-FL/FedLab}}
    
    \item SecureBoost \cite{Cheng2021SecureBoost} (and its SecureBoost+ variant) are designed for vertical federated learning with gradient-boosted decision trees which is optimal for feature-partitioned tabular data. This is incompatible with our horizontal, neural-network-based ICD classification scenario. 
    
\end{itemize}

We selected then Flower for this study due to its cross-platform compatibility, framework-free design, active development community and ease of setting up simulations across 20 nodes with PyTorch integration. 

\subsection{Experimental Setup}\label{sec:setup}

\subsubsection*{Experiments: Centralized \& Federated training}

Two experimental training paradigms in this study are considered. In the centralized training setting, models are trained on the full dataset using both the public stratified split proposed by \cite{nguyen2023mimic} and 10 additional stratified random splits for ablation. These experiments serve as baseline evaluations to benchmark the effectiveness of our embedding-based classification approach.

For the federated training setup, we simulate a horizontal FL environment using the Flower framework \cite{Beutel2020Flower}, which supports flexible prototyping and mixed-platform development. The full training data is randomly partitioned into 20 disjoint subsets, one per client, to simulate the non-deterministic and heterogeneous nature of real-world hospital configuration. Each client retains its local data and participates in collaborative training via a FedAvg-based aggregation strategy. This setup is configured for 30 communication rounds, with all clients participating in each training round (\texttt{fraction\_fit = 1.0}) and half participating in evaluation (\texttt{fraction\_evaluate = 0.5}). Minimum thresholds are set for client availability (\texttt{min\_available\_clients = 10}), fitting (\texttt{min\_fit\_clients = 10}) and evaluation (\texttt{min\_evaluate\_clients = 5}). Evaluation metrics are aggregated using a weighted average and training progress is tracked via weighted training loss, accounting for variation in local dataset sizes.

This experimental setup allows to compare between the centralized and federated learning under controlled conditions. It can also help to isolate the effect of federated constraints on performance without confounding factors such as inconsistent preprocessing, data scale, or model architecture.

\subsubsection*{Data preprocessing \& Splits}
To prevent data leakage and ensure proper generalization evaluation, feature normalization was performed strictly using the training set statistics. Specifically, embedding vectors were standardized with a \texttt{StandardScaler} fitted only on the training subset and subsequently applied to the validation and test sets. This design guarantees that no information from unseen data influenced the training process. Additionally, multi-label targets were binarized using \texttt{MultiLabelBinarizer}, fitted once on the full set of training labels. Both the normalization scaler and label binarizer were saved, ensuring consistency for future inference and evaluation on newly recorded data under identical preprocessing conditions. 

For baseline benchmarking, we adopted the stratified data split proposed in \cite{nguyen2023mimic} for MIMIC-VI discharge notes. This approach ensures that label distributions are approximately preserved across training, validation and test sets and provides a reproducible reference point for comparing with other methods. However, it is important to note that the original study focuses on centralized training, often employing large models trained from scratch to maximize performance. In contrast, our setup operates under FL constraints, where pretraining and data access are restricted on nodes only.
To further assess the robustness of our method, we conducted an ablation study using 10 additional independent stratified splits (train-validation-test ratio: 0.70, 0.15, 0.15). These experiments were performed under centralized training only, as our results (detailed in results section) showed that performance across centralized and FL settings was highly consistent when using embedding-based representations. Reported results for these additional splits include both mean and standard deviation.

\subsubsection*{Training parameters}

In this study, we used HybridLoss function combining two commonly used objectives for imbalanced multi-label classification: Binary Cross-Entropy (BCE) and Focal Loss. The loss is defined as:
$$\mathcal{L}_{\text{hybrid}} = \lambda_{\text{bce}} \cdot \mathcal{L}_{\text{BCE}} + (1 - \lambda_{\text{bce}}) \cdot \mathcal{L}_{\text{focal}}$$
where, the Binary Cross-Entropy loss and the Focal Loss are defined as:
\begin{align*}
\mathcal{L}_{\text{BCE}} &= - \sum_{i=1}^{C} \left[ y_i \log \hat{y}_i + (1 - y_i) \log (1 - \hat{y}_i) \right] \\
\mathcal{L}_{\text{focal}} &= - \sum_{i=1}^{C} \left[ 
\alpha (1 - \hat{y}_i)^\gamma y_i \log \hat{y}_i + 
(1 - \alpha) \hat{y}_i^\gamma (1 - y_i) \log (1 - \hat{y}_i) 
\right]
\end{align*}
Here, \( \hat{y}_i \) is the predicted probability for label \( i \), \( y_i \in \{0, 1\} \) is the ground-truth label, \( \alpha \in [0, 1] \) controls class imbalance, \( \gamma \geq 0 \) focuses learning on hard examples, \( C \) is the total number of labels and \( \lambda_{\text{bce}} \in [0,1] \) balances the contribution of the two components.

In our experiments, we used $\alpha=0.35$, $\gamma=2$, and $\lambda_{\text{bce}}=0.5$ to balance sensitivity to rare labels and training stability. All models were trained using the \texttt{AdamW} optimizer with a learning rate of $10^{-3}$, a weight decay of $10^{-5}$, and a cosine annealing learning rate scheduler over 20 epochs. A dropout rate of 0.1 was used in all layers to mitigate overfitting, especially given the modest model sizes and high label dimensionality. 

To ensure clean isolation and avoid backend instability, especially when using the Ray engine in Flower, each training session is executed within a fresh Python kernel. This design decision mitigates resource leakage and prevents cumulative errors across multiple runs, which we observed especially under Windows environments.

\section{Results and Insights}\label{sec:results}

Unlike most prior work in automated ICD classification where model evaluation is often based on a single training instance with fixed hyperparameters and splits, this study adopts a more exhaustive evaluation protocol. For each of the six selected embedding models, we trained three classifier variants (MLP, DeepMLP and DeepResMLP) under two configurations: centralized training and federated learning (FL) with 20 simulated clients. 
Additionally, to assess robustness under different data splits, we conducted an ablation study using 10 independent random stratified splits in the centralized setting. In total, this results in $6 \times 3 \times 2 + 6 \times 3 \times 10 = 216$ core training runs used for the reported results, with a further $6 \times 3 \times 2 = 36$ exploratory long runs at 100 communication rounds that did not improve performance. Taken together, these experiments offer a broader and more statistically grounded view of performance variance than previously published benchmarks.

Recent efforts in automated ICD coding have focused on improving both model accuracy and evaluation rigor. A particularly relevant contribution is the comprehensive benchmark study by Edin et al. \cite{Edin2023Automated}, which re-implemented and compared SOTA models on MIMIC-III and MIMIC-IVs. While their setup permits full access to training data and end-to-end fine-tuning, our design prioritizes data locality and modularity, so the comparison is only partially aligned. Still, the relative performance observed here provides a compelling argument that privacy-respecting frameworks with modular embedding pipelines can yield competitive results without sacrificing regulatory compliance.

\subsubsection*{Main results}

\begin{table}[!h]
\footnotesize
\centering
\caption{Performance of centralized training across six embedding models and three classifier variants (MLP, DeepMLP, DeepResMLP) for ICD-10 and ICD-9 tasks on MIMIC-IV. Best scores per metric are highlighted in gray. Results are reported using macro and micro-averaged precision, recall, and F1 scores.}
\label{tab:res_central}
\begin{tabular}{ll|C{1.5cm}C{1cm}C{1cm}|C{1.5cm}C{1cm}C{1cm}}
\toprule
\multicolumn{8}{c}{\textbf{MIMIC IV - ICD 10}} \\ \midrule
Embedding & Model & Macro Precision & Macro Recall & Macro F1 & Micro Precision & Micro Recall & Micro F1 \\
\midrule
\qwenthree & \mlp & \cellcolor{gray!40}0.2726 & 0.2919 & 0.2685 & \cellcolor{gray!40}0.3435 & 0.4625 & 0.3942 \\
\qwenthree & \deepmlp & 0.2090 & 0.3034 & 0.2255 & 0.2790 & 0.5272 & 0.3649 \\
\qwenthree & \deepresmlp & 0.2470 & 0.3270 & 0.2670 & 0.3084 & 0.5195 & 0.3870 \\
\qwentwo & \mlp & 0.2568 & 0.2975 & 0.2660 & 0.3396 & 0.4760 & \cellcolor{gray!40}0.3964 \\
\qwentwo & \deepmlp & 0.2029 & 0.3039 & 0.2229 & 0.2830 & 0.5372 & 0.3707 \\
\qwentwo & \deepresmlp & 0.2391 & \cellcolor{gray!40}0.3423 & \cellcolor{gray!40}0.2709 & 0.3104 & \cellcolor{gray!40}0.5420 & 0.3947 \\
\gtemultilingual & \mlp & 0.2681 & 0.2814 & 0.2589 & 0.3384 & 0.4517 & 0.3869 \\
\gtemultilingual & \deepmlp & 0.2030 & 0.2912 & 0.2169 & 0.2732 & 0.5191 & 0.3580 \\
\gtemultilingual & \deepresmlp & 0.2387 & 0.3252 & 0.2605 & 0.2966 & 0.5246 & 0.3789 \\
\snowflakel & \mlp & 0.2242 & 0.3107 & 0.2484 & 0.2983 & 0.5156 & 0.3779 \\
\snowflakel & \deepmlp & 0.1932 & 0.2714 & 0.2048 & 0.2884 & 0.5115 & 0.3688 \\
\snowflakel & \deepresmlp & 0.2263 & 0.3152 & 0.2500 & 0.3036 & 0.5309 & 0.3863 \\
\snowflakem & \mlp & 0.2256 & 0.3126 & 0.2496 & 0.2893 & 0.5141 & 0.3702 \\
\snowflakem & \deepmlp & 0.1976 & 0.2748 & 0.2081 & 0.2819 & 0.5079 & 0.3626 \\
\snowflakem & \deepresmlp & 0.2314 & 0.3136 & 0.2507 & 0.2987 & 0.5224 & 0.3801 \\
\nomic & \mlp & 0.2157 & 0.2834 & 0.2295 & 0.2810 & 0.4965 & 0.3589 \\
\nomic & \deepmlp & 0.1799 & 0.2559 & 0.1898 & 0.2656 & 0.5002 & 0.3469 \\
\nomic & \deepresmlp & 0.2108 & 0.2914 & 0.2285 & 0.2816 & 0.5105 & 0.3630 \\
\toprule
\multicolumn{8}{c}{\textbf{MIMIC IV - ICD 9}} \\ \midrule
\qwenthree & \mlp & 0.2929 & 0.3081 & 0.2871 & 0.3662 & 0.4747 & 0.4135 \\
\qwenthree & \deepmlp & 0.2286 & 0.3304 & 0.2508 & 0.2888 & 0.5398 & 0.3763 \\
\qwenthree & \deepresmlp & 0.2690 & \cellcolor{gray!40}0.3492 & 0.2893 & 0.3220 & 0.5350 & 0.4020 \\
\qwentwo & \mlp & 0.2790 & 0.3086 & 0.2842 & 0.3670 & 0.4839 & 0.4174 \\
\qwentwo & \deepmlp & 0.2340 & 0.3184 & 0.2496 & 0.3123 & 0.5342 & 0.3942 \\
\qwentwo & \deepresmlp & \cellcolor{gray!40}0.2948 & 0.3183 & \cellcolor{gray!40}0.2934 & \cellcolor{gray!40}0.3806 & 0.5031 & \cellcolor{gray!40}0.4334 \\
\gtemultilingual & \mlp & 0.2787 & 0.2971 & 0.2746 & 0.3519 & 0.4639 & 0.4002 \\
\gtemultilingual & \deepmlp & 0.2245 & 0.3163 & 0.2422 & 0.2859 & 0.5258 & 0.3704 \\
\gtemultilingual & \deepresmlp & 0.2605 & 0.3369 & 0.2794 & 0.3163 & 0.5274 & 0.3954 \\
\snowflakel & \mlp & 0.2577 & 0.2922 & 0.2636 & 0.3513 & 0.4835 & 0.4069 \\
\snowflakel & \deepmlp & 0.2071 & 0.3070 & 0.2297 & 0.2910 & 0.5347 & 0.3769 \\
\snowflakel & \deepresmlp & 0.2498 & 0.3235 & 0.2683 & 0.3268 & 0.5308 & 0.4045 \\
\snowflakem & \mlp & 0.2665 & 0.2950 & 0.2674 & 0.3469 & 0.4786 & 0.4023 \\
\snowflakem & \deepmlp & 0.2168 & 0.3061 & 0.2337 & 0.2898 & 0.5278 & 0.3741 \\
\snowflakem & \deepresmlp & 0.2500 & 0.3351 & 0.2721 & 0.3158 & \cellcolor{gray!40}0.5400 & 0.3985 \\
\nomic & \mlp & 0.2325 & 0.3027 & 0.2494 & 0.2947 & 0.5068 & 0.3727 \\
\nomic & \deepmlp & 0.2012 & 0.2907 & 0.2174 & 0.2711 & 0.5196 & 0.3563 \\
\nomic & \deepresmlp & 0.2320 & 0.3171 & 0.2537 & 0.2928 & 0.5316 & 0.3776 \\
\bottomrule
\end{tabular}
\end{table}

Table~\ref{tab:res_central} summarizes the performance of all centralized training configurations. Results confirm that even with relatively lightweight classifiers, substantial variation in performance arises from the choice of embedding model. Among all configurations, \texttt{Qwen3-Embed-0.6B} with MLP achieves the highest micro F1 score on ICD-10 ($0.3942$), while \texttt{gte-Qwen2-1.5B-instruct} with DeepResMLP performs best on ICD-9 across all metrics (macro F1: $0.2934$; micro F1: $0.4334$). Interestingly, deeper classifiers like DeepResMLP did not consistently outperform MLP, which reinforces the core design hypothesis of this work: when semantically rich embeddings are used as input, the added complexity of deeper networks yields diminishing returns. This is particularly evident in configurations where MLP matches or even exceeds the DeepMLP/DeepResMLP performance. However, it is worth noting that the occasional underperformance of DeepResMLP may due to the data limitations rather than architectural inefficacy: with more training examples per label or richer training regimes, deeper networks might demonstrate clearer advantages. Therefore, we do not rule out the potential of these architectures in future extensions of this work.

\begin{table}[!t]
\footnotesize
\centering
\caption{Performance of federated training across six embedding models and three classifier variants (MLP, DeepMLP, DeepResMLP) for ICD-10 and ICD-9 tasks on MIMIC-IV. Best scores per metric are highlighted in gray. Results are reported using macro and micro-averaged precision, recall, and F1 scores.}
\label{tab:res_flwr}
\begin{tabular}{ll|C{1.5cm}C{1cm}C{1cm}|C{1.5cm}C{1cm}C{1cm}}
\toprule
\multicolumn{8}{c}{\textbf{MIMIC IV - ICD 10}} \\ \midrule
Embedding & Model & Macro Precision & Macro Recall & Macro F1 & Micro Precision & Micro Recall & Micro F1 \\
\midrule
\qwenthree & \mlp & \cellcolor{gray!40}0.2671 & 0.2917 & 0.2555 & \cellcolor{gray!40}0.3340 & 0.4816 & 0.3944 \\
\qwenthree & \deepmlp & 0.1841 & 0.2633 & 0.1934 & 0.2744 & 0.4973 & 0.3537 \\
\qwenthree & \deepresmlp & 0.2566 & 0.2876 & 0.2509 & 0.3185 & 0.4951 & 0.3876 \\
\qwentwo & \mlp & 0.2475 & 0.3160 & \cellcolor{gray!40}0.2616 & 0.3297 & 0.5040 & \cellcolor{gray!40}0.3986 \\
\qwentwo & \deepmlp & 0.1839 & 0.2633 & 0.1926 & 0.2813 & 0.5071 & 0.3618 \\
\qwentwo & \deepresmlp & 0.2397 & \cellcolor{gray!40}0.3303 & 0.2565 & 0.2728 & \cellcolor{gray!40}0.5740 & 0.3699 \\
\gtemultilingual & \mlp & 0.2221 & 0.3269 & 0.2450 & 0.2696 & 0.5386 & 0.3593 \\
\gtemultilingual & \deepmlp & 0.1819 & 0.2573 & 0.1901 & 0.2713 & 0.4906 & 0.3494 \\
\gtemultilingual & \deepresmlp & 0.2287 & 0.3238 & 0.2466 & 0.2544 & 0.5585 & 0.3496 \\
\snowflakel & \mlp & 0.2383 & 0.2817 & 0.2378 & 0.3294 & 0.4883 & 0.3934 \\
\snowflakel & \deepmlp & 0.1692 & 0.2447 & 0.1778 & 0.2771 & 0.4954 & 0.3554 \\
\snowflakel & \deepresmlp & 0.2255 & 0.2965 & 0.2312 & 0.2692 & 0.5532 & 0.3622 \\
\snowflakem & \mlp & 0.2088 & 0.3214 & 0.2339 & 0.2690 & 0.5405 & 0.3592 \\
\snowflakem & \deepmlp & 0.1698 & 0.2518 & 0.1798 & 0.2681 & 0.4932 & 0.3474 \\
\snowflakem & \deepresmlp & 0.2178 & 0.2908 & 0.2237 & 0.2504 & 0.5502 & 0.3441 \\
\nomic & \mlp & 0.1922 & 0.2886 & 0.2089 & 0.2590 & 0.5005 & 0.3413 \\
\nomic & \deepmlp & 0.1298 & 0.2916 & 0.1579 & 0.1959 & 0.5670 & 0.2911 \\
\nomic & \deepresmlp & 0.1823 & 0.2321 & 0.1791 & 0.2252 & 0.5036 & 0.3112 \\
\toprule
\multicolumn{8}{c}{\textbf{MIMIC IV - ICD 9}} \\ \midrule
\qwenthree & \mlp & 0.2675 & 0.3243 & 0.2727 & 0.3279 & 0.5057 & 0.3978 \\
\qwenthree & \deepmlp & 0.2090 & 0.2687 & 0.2090 & 0.2879 & 0.4937 & 0.3637 \\
\qwenthree & \deepresmlp & 0.2942 & 0.3072 & 0.2776 & 0.3371 & 0.5082 & 0.4053 \\
\qwentwo & \mlp & 0.2876 & 0.3112 & 0.2798 & \cellcolor{gray!40}0.3830 & 0.4750 & \cellcolor{gray!40}0.4241 \\
\qwentwo & \deepmlp & 0.2072 & 0.2694 & 0.2075 & 0.2928 & 0.5073 & 0.3713 \\
\qwentwo & \deepresmlp & \cellcolor{gray!40}0.2948 & 0.3141 & \cellcolor{gray!40}0.2836 & 0.3478 & 0.5270 & 0.4190 \\
\gtemultilingual & \mlp & 0.2526 & 0.3050 & 0.2570 & 0.3210 & 0.4913 & 0.3883 \\
\gtemultilingual & \deepmlp & 0.1992 & 0.2617 & 0.2010 & 0.2793 & 0.4883 & 0.3553 \\
\gtemultilingual & \deepresmlp & 0.2820 & 0.2999 & 0.2677 & 0.3249 & 0.5027 & 0.3947 \\
\snowflakel & \mlp & 0.2409 & 0.3008 & 0.2480 & 0.3317 & 0.5003 & 0.3989 \\
\snowflakel & \deepmlp & 0.1886 & 0.2466 & 0.1897 & 0.2885 & 0.4899 & 0.3632 \\
\snowflakel & \deepresmlp & 0.2318 & \cellcolor{gray!40}0.3286 & 0.2541 & 0.2760 & \cellcolor{gray!40}0.5754 & 0.3731 \\
\snowflakem & \mlp & 0.2527 & 0.2914 & 0.2483 & 0.3344 & 0.4876 & 0.3967 \\
\snowflakem & \deepmlp & 0.1926 & 0.2509 & 0.1920 & 0.2831 & 0.4881 & 0.3583 \\
\snowflakem & \deepresmlp & 0.2728 & 0.2819 & 0.2550 & 0.3315 & 0.5018 & 0.3993 \\
\nomic & \mlp & 0.2225 & 0.2657 & 0.2224 & 0.3245 & 0.4434 & 0.3747 \\
\nomic & \deepmlp & 0.1450 & 0.2935 & 0.1712 & 0.2046 & 0.5633 & 0.3001 \\
\nomic & \deepresmlp & 0.1712 & 0.1643 & 0.1426 & 0.2317 & 0.3895 & 0.2905 \\
\bottomrule
\end{tabular}
\end{table}

Across embedding models, \texttt{Qwen3-Embed-0.6B} and \qwentwo\, consistently achieve top-tier performance, particularly on the more fine-grained ICD-9 task. Despite being smaller (0.6B parameters), Qwen3-Embed performs competitively or better than larger models like \qwentwo\,, especially when paired with simple MLP classifiers. This suggests that model size is not the only driver of embedding quality in this context, alignment and training objective may play more essential roles. Conversely, \nomic\,, while much more efficient computationally, generally lags behind in performance metrics, possibly due to domain mismatch or shorter context window optimization. The Snowflake-Arctic models offer a solid middle ground in terms of both speed and predictive quality. 

Comparing both ICD versions, ICD-9 classification results tend to slightly outperform ICD-10. In this study, 207\,817 discharge notes labeled with ICD-9 codes and 121\,344 notes with ICD-10 codes were considered. Although ICD-10 is the more recent and granular standard, the observed data imbalance stems from the label distribution of the MIMIC-IV dataset. Specifically, ICD-9 codes dominate the time span of available records (2008–2015), whereas ICD-10 appears only in later admissions (post-2015). Additionally, the higher specificity of ICD-10 leads to a sparser label distribution. Combined with our requirement of at least 200 samples per label to support federated partitioning across 20 clients, these factors explain the reduced volume of ICD-10 examples relative to ICD-9 in the filtered dataset.

Table~\ref{tab:res_flwr} reports the performance of federated learning (FL) configurations using the same test data as in the centralized setting. Though the constraints imposed by data decentralization, most FL results closely match their centralized counterparts, especially for MLP classifiers, which continue to show surprisingly strong performance when paired with top-tier embedding models. For instance, \qwenthree\, + MLP achieves a micro F1 of $0.3944$ on ICD-10, nearly identical to centralized training ($0.3942$), highlighting that embedding-based FL workflows can retain classification accuracy even with strict data isolation.

\begin{table}[!htb]
\footnotesize
\centering
\caption{Performance of centralized training for the TOP10 \& TOP 50 frequent ICD codes.} 
\label{tab:res_central_TopK} 
\begin{tabular}{ll|ccc|ccc}
\toprule
\multicolumn{8}{c}{\textbf{MIMIC IV - ICD 10}} \\ \midrule
\multirow{2}{*}{\textbf{Embedding}}  & \multirow{2}{*}{\textbf{Model}}  & \multicolumn{3}{c|}{\textbf{TOP 10}} & \multicolumn{3}{c}{\textbf{TOP 50}} \\
 &  & Precision & Recall & F1-score & Precision & Recall & F1-score \\ \midrule
\qwenthree & \mlp & 0.5425 & 0.6678 & 0.5893 & 0.4092 & 0.5123 & 0.4469 \\
\qwenthree & \deepmlp & 0.5171 & 0.6898 & 0.5867 & 0.3877 & 0.5331 & 0.4415 \\
\qwenthree & \deepresmlp & 0.5279 & 0.6963 & 0.5974 & 0.4088 & 0.5350 & 0.4555 \\
\qwentwo & \mlp & 0.5429 & 0.7131 & 0.6112 & 0.4202 & 0.5389 & 0.4653 \\
\qwentwo & \deepmlp & 0.5414 & 0.7076 & 0.6096 & 0.4115 & 0.5453 & 0.4615 \\
\qwentwo & \deepresmlp & \cellcolor{gray!40}0.5673 & 0.7042 & 0.6232 & \cellcolor{gray!40}0.4379 & 0.5475 & \cellcolor{gray!40}0.4796 \\
\gtemultilingual & \mlp & 0.5020 & 0.7031 & 0.5800 & 0.3768 & 0.5207 & 0.4289 \\
\gtemultilingual & \deepmlp & 0.4971 & 0.7053 & 0.5788 & 0.3716 & 0.5267 & 0.4279 \\
\gtemultilingual & \deepresmlp & 0.5250 & 0.6822 & 0.5900 & 0.3940 & 0.5248 & 0.4437 \\
\snowflakel & \mlp & 0.5466 & \cellcolor{gray!40}0.7221 & 0.6175 & 0.4128 & 0.5527 & 0.4654 \\
\snowflakel & \deepmlp & 0.5365 & 0.7170 & 0.6109 & 0.4058 & 0.5402 & 0.4572 \\
\snowflakel & \deepresmlp & 0.5593 & 0.7157 & \cellcolor{gray!40}0.6243 & 0.4262 & \cellcolor{gray!40}0.5565 & 0.4757 \\
\snowflakem & \mlp & 0.5415 & 0.6816 & 0.5989 & 0.3944 & 0.5261 & 0.4433 \\
\snowflakem & \deepmlp & 0.5311 & 0.6848 & 0.5945 & 0.3862 & 0.5294 & 0.4377 \\
\snowflakem & \deepresmlp & 0.5413 & 0.7046 & 0.6094 & 0.4022 & 0.5510 & 0.4568 \\
\nomic & \mlp & 0.5046 & 0.6850 & 0.5774 & 0.3770 & 0.5126 & 0.4253 \\
\nomic & \deepmlp & 0.4960 & 0.6979 & 0.5739 & 0.3621 & 0.5262 & 0.4209 \\
\nomic & \deepresmlp & 0.5216 & 0.6876 & 0.5895 & 0.3879 & 0.5160 & 0.4362 \\
\toprule
\multicolumn{8}{c}{\textbf{MIMIC IV - ICD 9}} \\ \midrule
\qwenthree & \mlp & 0.5594 & 0.6635 & 0.5973 & 0.4464 & 0.5127 & 0.4686 \\
\qwenthree & \deepmlp & 0.5413 & 0.6725 & 0.5924 & 0.4239 & 0.5214 & 0.4575 \\
\qwenthree & \deepresmlp & 0.5611 & 0.6725 & 0.6037 & 0.4555 & 0.5207 & 0.4764 \\
\qwentwo & \mlp & 0.5750 & 0.6606 & 0.6096 & 0.4537 & 0.5359 & 0.4853 \\
\qwentwo & \deepmlp & 0.5439 & \cellcolor{gray!40}0.7027 & 0.6062 & 0.4318 & 0.5477 & 0.4764 \\
\qwentwo & \deepresmlp & \cellcolor{gray!40}0.5762 & 0.6987 & \cellcolor{gray!40}0.6238 & \cellcolor{gray!40}0.4712 & \cellcolor{gray!40}0.5531 & \cellcolor{gray!40}0.5030 \\
\gtemultilingual & \mlp & 0.5211 & 0.6599 & 0.5759 & 0.4056 & 0.5170 & 0.4492 \\
\gtemultilingual & \deepmlp & 0.5177 & 0.6604 & 0.5736 & 0.3954 & 0.5166 & 0.4417 \\
\gtemultilingual & \deepresmlp & 0.5361 & 0.6638 & 0.5851 & 0.4198 & 0.5243 & 0.4600 \\
\snowflakel & \mlp & 0.5616 & 0.6788 & 0.6069 & 0.4346 & 0.5495 & 0.4765 \\
\snowflakel & \deepmlp & 0.5392 & 0.6864 & 0.6002 & 0.4204 & 0.5443 & 0.4672 \\
\snowflakel & \deepresmlp & 0.5623 & 0.6905 & 0.6155 & 0.4467 & 0.5531 & 0.4873 \\
\snowflakem & \mlp & 0.5411 & 0.6726 & 0.5932 & 0.4187 & 0.5322 & 0.4608 \\
\snowflakem & \deepmlp & 0.5361 & 0.6669 & 0.5889 & 0.4039 & 0.5309 & 0.4506 \\
\snowflakem & \deepresmlp & 0.5460 & 0.6948 & 0.6041 & 0.4289 & 0.5447 & 0.4734 \\
\nomic & \mlp & 0.5203 & 0.6446 & 0.5700 & 0.3986 & 0.5059 & 0.4390 \\
\nomic & \deepmlp & 0.5042 & 0.6602 & 0.5658 & 0.3895 & 0.5032 & 0.4305 \\
\nomic & \deepresmlp & 0.5297 & 0.6542 & 0.5806 & 0.4102 & 0.5251 & 0.4534 \\
\bottomrule
\end{tabular}
\end{table}

On ICD-9, \texttt{gte-Qwen2-1.5B-instruct} + MLP maintains the best micro F1 ($0.4241$) and comparable macro metrics, again echoing its centralized dominance. Interestingly, deeper classifiers (e.g., DeepResMLP) underperform more variably across configurations. This may stem from convergence sensitivity in federated setups with more model parameters and asynchronous weight averaging strategy. Yet, DeepResMLP with \texttt{gte-Qwen2-1.5B-instruct} remains strong (macro F1: $0.2836$, micro F1: $0.4190$), confirming robustness under FL constraints.

Taken together, these findings reinforce the central hypothesis: modular FL systems can effectively inherit the semantic capacity of LLM embeddings without centralizing sensitive data. The minimal performance drop between centralized and FL variants, especially on macro F1, also underscores the stability of embedding-based representations across decentralized nodes.

Given the extreme multi-label nature of ICD-10 or ICD-9 code classification with over a thousand unique codes, macro and micro F1 scores are used as primary evaluation metrics, as they respectively reflect label-level balance and overall instance-level performance. Accuracy, or even label-based precision alone, may be misleading in these highly sparse and skewed label distributions.

\begin{table}[!t]
\footnotesize
\centering
\caption{Performance of federated training for the TOP10 \& TOP 50 frequent ICD codes.}
\label{tab:res_flwr_TopK} 
\begin{tabular}{ll|ccc|ccc}
\toprule
\multicolumn{8}{c}{\textbf{MIMIC IV - ICD 10}} \\ \midrule
\multirow{2}{*}{\textbf{Embedding}}  & \multirow{2}{*}{\textbf{Model}}  & \multicolumn{3}{c|}{\textbf{TOP 10}} & \multicolumn{3}{c}{\textbf{TOP 50}} \\
 &  & Precision & Recall & F1-score & Precision & Recall & F1-score \\ \midrule
\qwenthree & \mlp & 0.5196 & 0.6889 & 0.5874 & 0.3987 & 0.5116 & 0.4409 \\
\qwenthree & \deepmlp & 0.4965 & 0.6859 & 0.5723 & 0.3597 & 0.5190 & 0.4179 \\
\qwenthree & \deepresmlp & 0.5356 & 0.6683 & 0.5885 & 0.3976 & 0.5195 & 0.4429 \\
\qwentwo & \mlp & 0.5378 & 0.7186 & 0.6091 & 0.4175 & 0.5439 & 0.4645 \\
\qwentwo & \deepmlp & 0.5072 & \cellcolor{gray!40}0.7372 & 0.5956 & 0.3805 & 0.5322 & 0.4384 \\
\qwentwo & \deepresmlp & 0.5429 & 0.7173 & 0.6141 & \cellcolor{gray!40}0.4238 & 0.5436 & \cellcolor{gray!40}0.4681 \\
\gtemultilingual & \mlp & 0.4929 & 0.7132 & 0.5790 & 0.3725 & 0.5226 & 0.4279 \\
\gtemultilingual & \deepmlp & 0.4799 & 0.7084 & 0.5649 & 0.3430 & 0.5227 & 0.4066 \\
\gtemultilingual & \deepresmlp & 0.5005 & 0.7037 & 0.5813 & 0.3754 & 0.5299 & 0.4314 \\
\snowflakel & \mlp & \cellcolor{gray!40}0.5591 & 0.6969 & \cellcolor{gray!40}0.6168 & 0.4135 & 0.5414 & 0.4623 \\
\snowflakel & \deepmlp & 0.5318 & 0.6965 & 0.5988 & 0.3801 & 0.5245 & 0.4344 \\
\snowflakel & \deepresmlp & 0.5539 & 0.7023 & 0.6142 & 0.4099 & \cellcolor{gray!40}0.5483 & 0.4618 \\
\snowflakem & \mlp & 0.5219 & 0.7074 & 0.5952 & 0.3820 & 0.5324 & 0.4372 \\
\snowflakem & \deepmlp & 0.4977 & 0.7073 & 0.5807 & 0.3548 & 0.5209 & 0.4153 \\
\snowflakem & \deepresmlp & 0.5318 & 0.6904 & 0.5952 & 0.3862 & 0.5350 & 0.4392 \\
\nomic & \mlp & 0.4847 & 0.6896 & 0.5608 & 0.3539 & 0.5087 & 0.4087 \\
\nomic & \deepmlp & 0.4633 & 0.6988 & 0.5534 & 0.3255 & 0.5195 & 0.3931 \\
\nomic & \deepresmlp & 0.4812 & 0.6697 & 0.5551 & 0.3356 & 0.5117 & 0.3971 \\
\toprule
\multicolumn{8}{c}{\textbf{MIMIC IV - ICD 9}} \\ \midrule
\qwenthree & \mlp & 0.5400 & 0.6643 & 0.5881 & 0.4313 & 0.5025 & 0.4573 \\
\qwenthree & \deepmlp & 0.5012 & 0.6763 & 0.5671 & 0.3821 & 0.5097 & 0.4259 \\
\qwenthree & \deepresmlp & 0.5510 & 0.6650 & 0.5947 & 0.4348 & 0.5219 & 0.4646 \\
\qwentwo & \mlp & 0.5468 & 0.6921 & 0.6048 & 0.4386 & \cellcolor{gray!40}0.5523 & 0.4828 \\
\qwentwo & \deepmlp & 0.5175 & 0.6894 & 0.5846 & 0.3969 & 0.5346 & 0.4473 \\
\qwentwo & \deepresmlp & \cellcolor{gray!40}0.5596 & \cellcolor{gray!40}0.7045 & \cellcolor{gray!40}0.6128 & \cellcolor{gray!40}0.4552 & 0.5506 & \cellcolor{gray!40}0.4917 \\
\gtemultilingual & \mlp & 0.5035 & 0.6831 & 0.5712 & 0.3889 & 0.5215 & 0.4385 \\
\gtemultilingual & \deepmlp & 0.4798 & 0.6817 & 0.5525 & 0.3587 & 0.5140 & 0.4124 \\
\gtemultilingual & \deepresmlp & 0.5172 & 0.6684 & 0.5748 & 0.4074 & 0.5161 & 0.4480 \\
\snowflakel & \mlp & 0.5388 & 0.6834 & 0.5971 & 0.4200 & 0.5437 & 0.4669 \\
\snowflakel & \deepmlp & 0.5059 & 0.6808 & 0.5761 & 0.3820 & 0.5294 & 0.4362 \\
\snowflakel & \deepresmlp & 0.5545 & 0.6714 & 0.6035 & 0.4332 & 0.5455 & 0.4769 \\
\snowflakem & \mlp & 0.5178 & 0.6852 & 0.5835 & 0.4044 & 0.5287 & 0.4521 \\
\snowflakem & \deepmlp & 0.5003 & 0.6629 & 0.5651 & 0.3691 & 0.5159 & 0.4211 \\
\snowflakem & \deepresmlp & 0.5411 & 0.6675 & 0.5920 & 0.4136 & 0.5335 & 0.4593 \\
\nomic & \mlp & 0.4953 & 0.6574 & 0.5545 & 0.3749 & 0.5011 & 0.4211 \\
\nomic & \deepmlp & 0.4731 & 0.6609 & 0.5423 & 0.3384 & 0.5067 & 0.3965 \\
\nomic & \deepresmlp & 0.4520 & 0.6285 & 0.5164 & 0.3233 & 0.4581 & 0.3701 \\
\bottomrule
\end{tabular}
\end{table}

\subsubsection*{Results in Top 10 \& Top 50 frequent codes}

To provide additional insight beyond macro and micro F1, especially in case of extreme multi-label classification settings with long-tail distributions, we evaluated also the performance on the most frequent labels. Tables~\ref{tab:res_central_TopK} and~\ref{tab:res_flwr_TopK} report precision, recall and F1 scores over the top 10 and top 50 most common ICD codes for both centralized and federated configurations. 
Across both ICD-9 and ICD-10 datasets, we observed that performance is substantially higher in the Top-K setting, as expected due to higher support per label. F1 scores on Top-10 often exceed $0.60$, compared to macro F1s near or below $0.30$. \qwentwo\, and \snowflakel, particularly when paired with DeepResMLP, consistently showed top performance, achieving Top-10 F1s above $0.62$ and Top-50 F1s close to $0.50$. These results suggest strong discriminative capacity in frequent-label data. Again, federated training closely matches centralized results, with negligible degradation on Top-K labels. This further validated the embedding-based strategy's robustness under data distribution fragmentation. Interestingly, MLP often performs nearly as well as or better than deeper models (e.g., DeepMLP), especially in Top-10 performance, reinforcing the study's hypothesis: rich embeddings can obviate the need for deep classifiers, particularly on well-represented labels. On the ICD-9 setting, DeepResMLP with \qwentwo\, achieves the highest F1 on both Top-10 ($0.6238$ centralized, $0.6128$ FL) and Top-50 ($0.5030$ centralized, $0.4917$ FL), illustrating strong model-label alignment.

\subsubsection*{Ablation study: random stratified splits}

\begin{table}[!h]
\renewcommand{\arraystretch}{1.2}
\setlength{\tabcolsep}{3pt}
\footnotesize
\centering
\caption{Performance of centralized training, 10 Stratified Random Split; results showed with mean $\pm$ standard deviation}
\label{tab:ablation_rsplits}
\begin{adjustbox}{angle=90,width=0.9\textwidth}
\begin{tabular}{ll|C{2.1cm}C{2.1cm}C{2.1cm}|C{2.1cm}C{2.1cm}C{2.1cm}}
\toprule
\multicolumn{8}{c}{\textbf{MIMIC IV - ICD 10}} \\ \midrule
Embedding & Model & Macro Precision & Macro Recall & Macro F1 & Micro Precision & Micro Recall & Micro F1 \\
\midrule
\qwenthree & \mlp & \cellcolor{gray!40}$0.2686 \pm 0.0141$ & $0.3032 \pm 0.0184$ & \cellcolor{gray!40}$0.2723 \pm 0.0015$ & $0.3337 \pm 0.0210$ & $0.4836 \pm 0.0249$ & $0.3938 \pm 0.0073$ \\
\qwenthree & \deepmlp & $0.2042 \pm 0.0022$ & $0.2767 \pm 0.0082$ & $0.2113 \pm 0.0049$ & $0.2811 \pm 0.0028$ & $0.5207 \pm 0.0059$ & $0.3651 \pm 0.0024$ \\
\qwenthree & \deepresmlp & $0.2467 \pm 0.0038$ & \cellcolor{gray!40}$0.3215 \pm 0.0054$ & $0.2663 \pm 0.0015$ & $0.3096 \pm 0.0052$ & $0.5211 \pm 0.0055$ & $0.3884 \pm 0.0028$ \\
\qwentwo & \mlp & $0.2662 \pm 0.0029$ & $0.2956 \pm 0.0032$ & $0.2721 \pm 0.0007$ & $0.3484 \pm 0.0051$ & $0.4762 \pm 0.0057$ & $0.4023 \pm 0.0018$ \\
\qwentwo & \deepmlp & $0.1996 \pm 0.0140$ & $0.2825 \pm 0.0087$ & $0.2109 \pm 0.0058$ & $0.2834 \pm 0.0175$ & \cellcolor{gray!40}$0.5372 \pm 0.0168$ & $0.3704 \pm 0.0128$ \\
\qwentwo & \deepresmlp & $0.2615 \pm 0.0126$ & $0.3071 \pm 0.0166$ & $0.2722 \pm 0.0010$ & \cellcolor{gray!40}$0.3492 \pm 0.0211$ & $0.5018 \pm 0.0232$ & \cellcolor{gray!40}$0.4108 \pm 0.0077$ \\
\gtemultilingual & \mlp & $0.2401 \pm 0.0123$ & $0.3125 \pm 0.0155$ & $0.2598 \pm 0.0010$ & $0.3008 \pm 0.0176$ & $0.5051 \pm 0.0213$ & $0.3763 \pm 0.0068$ \\
\gtemultilingual & \deepmlp & $0.1975 \pm 0.0061$ & $0.2679 \pm 0.0046$ & $0.2033 \pm 0.0045$ & $0.2746 \pm 0.0041$ & $0.5132 \pm 0.0037$ & $0.3577 \pm 0.0033$ \\
\gtemultilingual & \deepresmlp & $0.2356 \pm 0.0041$ & $0.3160 \pm 0.0081$ & $0.2575 \pm 0.0015$ & $0.2990 \pm 0.0057$ & $0.5218 \pm 0.0080$ & $0.3801 \pm 0.0029$ \\
\snowflakel & \mlp & $0.2347 \pm 0.0134$ & $0.3047 \pm 0.0177$ & $0.2552 \pm 0.0013$ & $0.3145 \pm 0.0205$ & $0.5109 \pm 0.0243$ & $0.3883 \pm 0.0078$ \\
\snowflakel & \deepmlp & $0.1843 \pm 0.0123$ & $0.2650 \pm 0.0123$ & $0.1951 \pm 0.0048$ & $0.2770 \pm 0.0188$ & $0.5288 \pm 0.0193$ & $0.3628 \pm 0.0139$ \\
\snowflakel & \deepresmlp & $0.2257 \pm 0.0039$ & $0.3106 \pm 0.0045$ & $0.2496 \pm 0.0018$ & $0.3065 \pm 0.0061$ & $0.5326 \pm 0.0058$ & $0.3890 \pm 0.0035$ \\
\snowflakem & \mlp & $0.2316 \pm 0.0034$ & $0.3132 \pm 0.0042$ & $0.2551 \pm 0.0011$ & $0.2967 \pm 0.0050$ & $0.5193 \pm 0.0052$ & $0.3776 \pm 0.0028$ \\
\snowflakem & \deepmlp & $0.1865 \pm 0.0134$ & $0.2675 \pm 0.0163$ & $0.1972 \pm 0.0048$ & $0.2713 \pm 0.0218$ & $0.5236 \pm 0.0252$ & $0.3563 \pm 0.0164$ \\
\snowflakem & \deepresmlp & $0.2318 \pm 0.0045$ & $0.3090 \pm 0.0081$ & $0.2517 \pm 0.0021$ & $0.3024 \pm 0.0049$ & $0.5255 \pm 0.0078$ & $0.3838 \pm 0.0022$ \\
\nomic & \mlp & $0.2130 \pm 0.0041$ & $0.2806 \pm 0.0055$ & $0.2275 \pm 0.0016$ & $0.2776 \pm 0.0061$ & $0.5046 \pm 0.0077$ & $0.3580 \pm 0.0032$ \\
\nomic & \deepmlp & $0.1670 \pm 0.0149$ & $0.2485 \pm 0.0244$ & $0.1763 \pm 0.0044$ & $0.2513 \pm 0.0246$ & $0.5135 \pm 0.0327$ & $0.3357 \pm 0.0179$ \\
\nomic & \deepresmlp & $0.2103 \pm 0.0057$ & $0.2801 \pm 0.0075$ & $0.2247 \pm 0.0028$ & $0.2822 \pm 0.0065$ & $0.5094 \pm 0.0086$ & $0.3631 \pm 0.0036$ \\
\toprule
\multicolumn{8}{c}{\textbf{MIMIC IV - ICD 9}} \\ \midrule
\qwenthree & \mlp & $0.3009 \pm 0.0038$ & $0.3118 \pm 0.0018$ & $0.2953 \pm 0.0012$ & $0.3699 \pm 0.0045$ & $0.4755 \pm 0.0034$ & $0.4161 \pm 0.0017$ \\
\qwenthree & \deepmlp & $0.2315 \pm 0.0034$ & $0.3182 \pm 0.0060$ & $0.2461 \pm 0.0014$ & $0.2936 \pm 0.0063$ & $0.5333 \pm 0.0068$ & $0.3786 \pm 0.0034$ \\
\qwenthree & \deepresmlp & \cellcolor{gray!40}$0.3038 \pm 0.0032$ & $0.3105 \pm 0.0033$ & $0.2930 \pm 0.0011$ & $0.3753 \pm 0.0048$ & $0.4829 \pm 0.0039$ & $0.4223 \pm 0.0016$ \\
\qwentwo & \mlp & $0.2848 \pm 0.0019$ & $0.3146 \pm 0.0010$ & $0.2925 \pm 0.0011$ & $0.3684 \pm 0.0027$ & $0.4868 \pm 0.0017$ & $0.4194 \pm 0.0013$ \\
\qwentwo & \deepmlp & $0.2281 \pm 0.0027$ & $0.3211 \pm 0.0034$ & $0.2464 \pm 0.0011$ & $0.3005 \pm 0.0032$ & \cellcolor{gray!40}$0.5474 \pm 0.0040$ & $0.3880 \pm 0.0017$ \\
\qwentwo & \deepresmlp & $0.2955 \pm 0.0042$ & $0.3209 \pm 0.0036$ & \cellcolor{gray!40}$0.2982 \pm 0.0013$ & \cellcolor{gray!40}$0.3842 \pm 0.0062$ & $0.5024 \pm 0.0052$ & \cellcolor{gray!40}$0.4354 \pm 0.0021$ \\
\gtemultilingual & \mlp & $0.2865 \pm 0.0039$ & $0.3017 \pm 0.0034$ & $0.2825 \pm 0.0009$ & $0.3556 \pm 0.0060$ & $0.4664 \pm 0.0055$ & $0.4034 \pm 0.0020$ \\
\gtemultilingual & \deepmlp & $0.2250 \pm 0.0034$ & $0.3104 \pm 0.0037$ & $0.2386 \pm 0.0014$ & $0.2846 \pm 0.0037$ & $0.5270 \pm 0.0046$ & $0.3696 \pm 0.0020$ \\
\gtemultilingual & \deepresmlp & $0.2789 \pm 0.0168$ & $0.3197 \pm 0.0185$ & $0.2834 \pm 0.0010$ & $0.3420 \pm 0.0246$ & $0.5008 \pm 0.0265$ & $0.4051 \pm 0.0092$ \\
\snowflakel & \mlp & $0.2730 \pm 0.0025$ & $0.2951 \pm 0.0029$ & $0.2747 \pm 0.0010$ & $0.3626 \pm 0.0042$ & $0.4799 \pm 0.0046$ & $0.4131 \pm 0.0012$ \\
\snowflakel & \deepmlp & $0.2137 \pm 0.0032$ & $0.2969 \pm 0.0039$ & $0.2271 \pm 0.0011$ & $0.2925 \pm 0.0041$ & $0.5330 \pm 0.0050$ & $0.3777 \pm 0.0021$ \\
\snowflakel & \deepresmlp & $0.2506 \pm 0.0028$ & \cellcolor{gray!40}$0.3282 \pm 0.0034$ & $0.2730 \pm 0.0010$ & $0.3264 \pm 0.0040$ & $0.5358 \pm 0.0041$ & $0.4056 \pm 0.0021$ \\
\snowflakem & \mlp & $0.2819 \pm 0.0042$ & $0.3013 \pm 0.0033$ & $0.2801 \pm 0.0010$ & $0.3578 \pm 0.0066$ & $0.4804 \pm 0.0059$ & $0.4101 \pm 0.0023$ \\
\snowflakem & \deepmlp & $0.2207 \pm 0.0025$ & $0.3035 \pm 0.0021$ & $0.2330 \pm 0.0008$ & $0.2892 \pm 0.0027$ & $0.5317 \pm 0.0031$ & $0.3747 \pm 0.0015$ \\
\snowflakem & \deepresmlp & $0.2659 \pm 0.0167$ & $0.3264 \pm 0.0187$ & $0.2793 \pm 0.0010$ & $0.3345 \pm 0.0245$ & $0.5248 \pm 0.0264$ & $0.4072 \pm 0.0092$ \\
\nomic & \mlp & $0.2371 \pm 0.0025$ & $0.3022 \pm 0.0026$ & $0.2534 \pm 0.0012$ & $0.2984 \pm 0.0038$ & $0.5069 \pm 0.0039$ & $0.3756 \pm 0.0020$ \\
\nomic & \deepmlp & $0.2012 \pm 0.0025$ & $0.2771 \pm 0.0028$ & $0.2108 \pm 0.0010$ & $0.2733 \pm 0.0029$ & $0.5126 \pm 0.0031$ & $0.3565 \pm 0.0018$ \\
\nomic & \deepresmlp & $0.2386 \pm 0.0046$ & $0.3051 \pm 0.0038$ & $0.2534 \pm 0.0013$ & $0.3034 \pm 0.0055$ & $0.5168 \pm 0.0059$ & $0.3823 \pm 0.0028$ \\
\bottomrule
\end{tabular}
\end{adjustbox}
\end{table}

To ensure robustness beyond a fixed benchmark split, we conducted an ablation study across 10 random stratified data partitions. Results (Table~\ref{tab:ablation_rsplits}) show that top-performing configurations maintain both high average performance and low variance, confirming the reliability of these findings.

The performance stability is generally high across most configurations. Standard deviations for F1 scores are often in the range of $0.001–0.003$, suggesting that the models are robust to variations in stratified train/test partitions. The \qwentwo\, + DeepResMLP combination emerges again as the most consistent top performer, achieving highest Micro F1 ($0.4108 \pm 0.0077$) and among the top Macro F1 ($0.2722 \pm 0.0010$) for ICD-10; and also high Micro F1 ($0.4354 \pm 0.0021$) and Macro F1 ($0.2982 \pm 0.0013$) for ICD-9 code, reinforcing prior findings from the federated and centralized setups. 
Interestingly, MLP models remain still competitive or even superior in Macro F1 in some configurations (e.g., \qwenthree\, + MLP in ICD-10), again highlighting that embedding quality can diminish the marginal benefit of classifier depth. DeepMLP models generally underperform, particularly in Macro F1 and occasionally show larger standard deviations, suggesting they may be less robust to training noise or sensitive to overfitting under certain embeddings.

Among embedding models, \qwentwo\, and \qwenthree\, consistently produce strong results across models and datasets. \nomic\,, in contrast, yields the lowest Macro/Micro F1 values regardless of classifier depth, reaffirming its relatively weaker discriminative capacity in this task. 

The higher performance on ICD-9 compared to ICD-10 is preserved across random splits, possibly reflecting more consistent labeling or higher support across commonly used ICD-9 codes, though the ICD-9 note set is also larger.

\section{Discussion}\label{sec:limits}

While recent benchmark studies such as \cite{Edin2023Automated} report SOTA performance using large-scale, end-to-end trained models like PLM-ICD (with approximately 139 million parameters), our results demonstrate that high-quality performance can also be achieved with drastically simpler and lighter-weight architectures. Specifically, while PLM-ICD reports micro F1 scores of 62.6\% and 58.5\% on MIMIC-IV ICD-9 and ICD-10 respectively, our best configurations—relying on frozen large language model (LLM) embeddings and shallow classifiers (MLP or DeepResMLP)—reach micro F1 scores of 43.5\% (ICD-9) and 41.1\% (ICD-10) and macro F1 scores of 29.8\% (ICD-9) and 27.2\% (ICD-10). These macro F1 scores are particularly noteworthy, as they approach or even match the PLM-ICD results, despite using models with less than 5\% of the parameter count and without any end-to-end training. Moreover, differences in label space, our models use 1\,085 and 1\,267 unique codes for ICD-10 and ICD-9 respectively, compared to the full codebooks likely used in the benchmark—may partly explain differences in micro F1. Importantly, our approach is more resource-efficient and adaptable, making it suited to real-world scenarios such as federated learning or low-resource clinical settings, where full-scale fine-tuning of large transformers is often less practical. These findings suggest that high-performing and modular medical coding systems can be built without relying on costly and rigid end-to-end architectures.

While this study demonstrates that frozen embedding models paired with lightweight MLP classifiers can achieve competitive results on multilabel ICD code classification tasks, several limitations are worth highlighting. First, the reliance on pretrained embeddings without task-specific fine-tuning may limit performance, particularly on domain-specific or rare codes. More expressive models such as those using attention, convolutional, or sequence architectures could potentially exploit the structure of clinical texts more effectively. A natural progression would be to explore a true end-to-end training pipeline, where embedding models are jointly fine-tuned with downstream classifiers. In addition, domain-specific LLMs that incorporate biomedical pretraining and hierarchical code structures could further improve robustness and generalization.

Another limitation is that we restricted our experiments to relatively simple classifier architectures, which may underfit more complex latent relationships in the data. Future work could explore hybrid architectures that integrate external medical knowledge, support hierarchical multi-label classification, or adapt label embeddings to reflect semantic code proximity. Moreover, our federated learning setup assumes idealized conditions, such as homogeneous data distribution across clients and consistent model configurations. Real-world FL systems would likely face greater variability in data volumes, label frequencies across nodes and client reliability, as well as communication inefficiencies and privacy constraints. Simulating more realistic federated environments-with heterogeneous data splits, asynchronous updates and differential privacy—would be a critical next step.

\section{Conclusion}\label{sec:concl}

This study explored the viability of federated learning (FL) for large-scale multi-label ICD code classification on clinical notes, with a focus on privacy-preserving training, standardized input representations and practical deployment scalability. Using frozen embedding models combined with lightweight multilayer perceptrons (MLPs), we demonstrated that competitive performance can be achieved without the need for fine-tuning large language models or centralizing sensitive patient data.

Through extensive experimentation, including centralized and federated setups, multiple embedding-classifier pairings and ablation studies over random data splits, we observed that embedding quality played a more critical role than classifier complexity. While deeper MLP variants occasionally offered marginal gains, simple MLPs performed robustly across configurations, reinforcing the notion that strong semantic representations can substantially reduce downstream model complexity. Moreover, our results showed that FL setups can yield performance comparable to centralized training under controlled conditions, suggesting promising paths for collaborative, privacy-compliant model development in healthcare network.

Nevertheless, several limitations remain, including the reliance on static embeddings, simplified FL assumptions and inherent label noise in medical coding datasets. Future work should investigate adaptive federated strategies, joint fine-tuning pipelines and domain-informed architectures. Importantly, addressing the ambiguity and inconsistency of ICD code annotations through uncertainty-aware evaluation or clinician feedback may be essential for real-world deployment. Overall, this study presents an encouraging step toward scalable and privacy-aware clinical NLP systems, while highlighting the trade-offs and design choices that must be investigated in federated healthcare AI.

\section*{ACKNOWLEDGMENTS}
{}This work is supported by the European Union's HORIZON Research and Innovation Programme, grant agreement No 101120657, project ENFIELD (European Lighthouse to Manifest Trustworthy and Green AI).

\bibliographystyle{ieeetr}
\bibliography{bib}

\end{document}